\documentclass[reprint,aps,showpacs,pra,nofootinbib]{revtex4-2}
\usepackage{newlfont}
\usepackage{braket,bbm}
\usepackage{undertilde}
\usepackage{amsmath,amsthm,mathrsfs}
\usepackage[varg]{txfonts}
\usepackage{xcolor,graphicx}

\DeclareMathAlphabet\mathcal{OMS}{cmsy}{m}{n}

\newcommand{\be}{\begin{equation}}
\newcommand{\ee}{\end{equation}}
\newcommand{\beq}{\begin{eqnarray}}
\newcommand{\eeq}{\end{eqnarray}}
\newcommand{\tx}[1]{\text{#1}}
\newcommand{\E}{\mathrm{e}}
\newcommand{\I}{\mathrm{i}}
\newcommand{\dr}{\text{d}}

\newcommand{\ms}[1]{\mathscr{#1}}
\newcommand{\mc}[1]{\mathcal{#1}}
\newcommand{\mf}[1]{\mathfrak{#1}}

\newcommand{\mbb}[1]{\mathbbm{#1}}
\newcommand{\vs}[2]{#1_{\mathrm{#2}}}
\newcommand{\ov}[1]{\overline{#1}}
\newcommand{\bpl}{\boldsymbol{(}}
\newcommand{\bpr}{\boldsymbol{)}}
\DeclareMathOperator{\Tr}{Tr}

\theoremstyle{definition}

\begin{document}
\title{Quantum resource covariance}
\author{M. F. Savi}
\email{msavi@fisica.ufpr.br}
\author{R. M. Angelo}
\email{renato@fisica.ufpr.br}
\affiliation{Department of Physics, Federal University of Paran\'a, P.O. Box 19044, 81531-980, Curitiba, Paran\'a, Brazil}

\begin{abstract}
Measurements invariably determine the physical state of a given ``observed system'' with respect to an ``observer system''. This implies an inescapable two-body relationality upon which physical theories must rely. There has been a growing effort to understand how quantum mechanics can accommodate the notion of reference frame in its substratum and whether such approach would respect general covariance. With regard to quantum resources, it has been shown that coherence and entanglement are not invariant upon changes of quantum reference frames. Here, we construct a theoretical framework within which a given combination of quantum resources, including not only coherence and entanglement, is shown to be a physical invariant. Under the premise of the universality of quantum phenomena, the proven covariance between quantum reference frames then implies that the total quantumness accessible to any observer in the universe is absolute.
\end{abstract} 

\maketitle

\section{Introduction}
Prompted by the theory of relativity, theoretical physics has inherited general covariance as one of its basic tenets. Such principle prescribes that coordinates are mere event markers upon which the ultimate laws of physics should not rely, that is, only relative positions matter. Crucially, this is a reliable ground for empirical science, for measurements are irreducibly relational probes of nature: they provide information about one system relative to another. It follows that a deep understanding of physical phenomena requires a careful account of the involved reference frames.

Even though classical physics has long ago induced us to conceive the idealization of absolute undisturbable frames of reference, there has been an increasing move toward recognizing reference frames as physical systems. Since the seminal demonstration that quantum mechanics can be consistently phrased from the viewpoint of a particle~\cite{aharonov1984}---a {\it quantum reference frame} (QRF)---the relevance of such construction has been emphasized in a variety of contexts involving fundamental aspects of physics~\cite{poulin2007,angelo2011,angelo2012,angelo2015},
superselection rules and quantum information~\cite{bartlett2006,bartlett2007}, quantum
communication~\cite{massar1995,bartlett2009}, entanglement detection~\cite{costa2009,liang2010}, resource theories~\cite{gour2008}, and thermodynamics~\cite{popescu2018}. The indispensable role of QRFs has shown to extend to relativistic quantum theory~\cite{peres2002,gingrich2002,peres2004,giacomini22019},
quantum gravity~\cite{rovelli1991-1,rovelli1991-2,dittrich2006,
dittrich2007,girelli2008,hohn2019,vanrietvelde2020,
ruiz2020}, cosmological models~\cite{bojowald2011-1,bojowald2011-2,hohn2012}, and quantum field theory~\cite{henderson2020,barbado2020}.

A distinctive feature of quantum theory lies in  its capability to deal with information, a fundamental quantum resource~\cite{lan1961,ben1987,acosta2020} powering quantum computing~\cite{nielsen2000}, quantum cryptography~\cite{shenoy2017}, and quantum thermodynamics~\cite{anders2017}. The huge amount of conceptual and technological developments achieved so far reveals that information is, if not the whole, a significant part of the story that quantum mechanics can tell us about nature~\cite{dakic2011,masanes2011,chiribella2011,hohn2017-Q,hohn2017-PRA}. Within the quantum formalism, information\footnote{As expressed in Eq.~\eqref{I}, information is a notion complementary to ignorance, with the latter being quantified via the von Neumann entropy. Recently, within the context of a resource theory of informational nonequilibrium in thermodynamics~\cite{gour2015}, $I(\rho)$ has been called {\it nonuniformity}, a resource whose meaning can be associated with the potential for doing informational work in an erasure operation.} can be quantified as~\cite{horodecki2003}
\be 
I(\rho)=\ln{d}-S(\rho),
\label{I}
\ee 
where $d$ is the dimension of the Hilbert space on which $\rho$ acts and $S$ is the von Neumann entropy. Since $\rho$ lives in a coordinate-independent space, the concept of information automatically respects general covariance. The formal proof is trivial for some symmetry groups: if the reference-frame transformation $\rho\mapsto\rho'=T\rho T^\dag$ is implemented through some unitary operation $T$, then the unitary invariance of the von Neumann entropy ensures that $I(\rho)=I(\rho')$. Note that this does not directly follow for more general groups~\cite{vanrietvelde2018,anne2020}. Information is stored, manipulated, and communicated through physical devices, and, in practice, is accessed via frame-dependent actions. Moreover, information can be encoded and distributed through quantum resources~\cite{chitambar2019} such as entanglement~\cite{cerf1997,horodecki2005,horodecki2007,horodecki2009} and coherence~\cite{baumgratz2014,girolami2014,streltsov2015,streltsov2017}. 

Now, it has been shown that entanglement and coherence are not invariant upon changes of QRFs~\cite{angelo2011,angelo2012,giacomini2019}. This implies that distinct reference systems have access to different amounts of quantum resources, even though the information encoded in the quantum state is the same. We are then led to speculate about the existence of some amount of resource that would respect invariance and how it would relate to the total information in $\rho$. For instance, is it the case that different observers may not agree on the amount of entanglement and coherence but do agree on a combination of them, much like one has in special relativity, where different observers do not agree on space and time intervals but do agree on the combination $\dr \mathbf{r}^{\,2}-(c\,\dr t)^2=\dr s^2$? A satisfactory answer to this question would make an important point in favour of fully relational quantum descriptions of nature, thus promoting quantum resources covariance to the role of a fundamental principle permeating many fields of physics.  

This work is devoted to make this point. To prove resources covariance, we first have to show how to decompose the invariant information $I(\rho)$ in terms of the quantum resources accessible from each local frame. This is done by means of a frame-independent procedure aimed at destroying all the information encoded in $\rho$. As we move toward the complete information erasing, we find that coherence and correlations do not suffice to expand the informational content of the state. This unveils the role of a nonclassical aspect that has remained unexplored so far in the QRFs literature.

\section{Information decomposition}
The Lorentz invariant element $\dr s^2$ can only be experimentally accessed through frame-dependent measurements of $\dr \mathbf{r}^{\,2}$ and $\dr t^2$. Here we show that the quantum invariant $I(\rho)$ is likewise decomposable in its local-frame constituents.  To this end, we devise a measurement-oriented procedure through which one guarantees that the entire information encoded in $\rho$ is erased in all reference frames. This is so because measurements are {\it events}; they occur in every reference frame, although with respect to distinct relational observables.

Since the generalization of our approach to multipartite systems is straightforward, we restrict our analysis to the bipartite case for simplicity. Consider the state $\rho\in\mf{B}(\mc{H})$ of two quantum systems, A and B, prepared with informational content given by Eq.~\eqref{I} with respect to a QRF R, where $\mf{B}(\mc{H})$ is the set of bounded operators acting on the Hilbert space $\mc{H=H_\tx{A}\otimes H_\tx{B}}$ of dimension $d=\vs{d}{A}\vs{d}{B}$. Let $A=\sum_ia_iA_i$ be a discrete-spectrum nondegenerate observable acting on $\vs{\mc{H}}{A}$, with corresponding projectors $A_i=\ket{a_i}\bra{a_i}$. After a measurement of this observable, the state collapses to $A_i\otimes\rho_{\tx{B}|i}$, where $\rho_{\tx{B}|i}=\bra{a_i}\rho\ket{a_i}/p_i$ and $p_i=\Tr(A_i\otimes\vs{\mathbbm{1}}{B}\rho)$. If the outcome $a_i$ is not revealed, the post-measurement state becomes 
\be
\sum_i^{\vs{d}{A}} p_i A_i \otimes \rho_{\tx{B}|i}=\sum_i^{\vs{d}{A}} \left(A_i\otimes\vs{\mathbbm{1}}{B}\right) \rho \left(A_i\otimes\vs{\mathbbm{1}}{B}\right)\eqqcolon \Phi_A(\rho).
\label{Phi}
\ee 
$\Phi_A$ is a completely positive trace-preserving map which indicates that an unrevealed measurement of $A$ has been performed in the reference frame R. Called dephasing map in the quantum resource theory of coherence~\cite{streltsov2015,streltsov2017}, on bipartite states this operation removes both coherence (in the $A$ basis) and entanglement. Most importantly, $\Phi_A$ manifests itself here as a key tool for our purposes, since it allows us to build the well-known quantifiers of (i) quantum coherence~\cite{baumgratz2014},
\be 
C_A(\vs{\rho}{A})\coloneqq S \bpl \Phi_A(\vs{\rho}{A}) \bpr -S(\vs{\rho}{A}),
\label{C_A}
\ee 
(ii) one-way quantum discord~\cite{ollivier2001,henderson2001,rulli2011},
\be 
D_A(\rho)\coloneqq \vs{I}{A:B}(\rho)-\vs{I}{A:B} \bpl \Phi_A(\rho) \bpr,
\label{D_A}
\ee 
and (iii) symmetric quantum discord 
\be 
D_{AB}(\rho)\coloneqq \vs{I}{A:B}(\rho)-\vs{I}{A:B} \bpl \Phi_{AB}(\rho) \bpr,
\label{D_AB}
\ee 
where $\vs{I}{A:B}(\rho)=S(\vs{\rho}{A})+S(\vs{\rho}{B})-S(\rho)$ is the mutual information between A and B, $\vs{\rho}{A(B)}=\vs{\Tr}{B(A)}(\rho)$ are reduced states, $\Phi_{AB}(\rho)\equiv\Phi_A\Phi_B(\rho)=\Phi_{BA}$ is a joint local map, and $\Phi_B$ is a map associated with observable $B=\sum_jb_jB_j\in\mf{B}(\mc{H}_\tx{B})$. Being basis dependent, the above measures are henceforth referred to as $A$-coherence, $A$-discord, and $AB$-discord, respectively  (similarly for measures related to observables on $\vs{\mc{H}}{B}$). Given the above, one checks that, upon a measurement of $A$, the implied informational decrease
\be 
I \bpl \Phi_A(\rho) \bpr -I(\rho)=-\big[C_A(\vs{\rho}{A})+D_A(\rho)\big]
\label{step1}
\ee 
corresponds to the amount of $A$-coherence and $A$-discord that are removed from $\rho$. Via direct calculations, we verify that $C_A \bpl \Phi_A(\rho) \bpr =D_A \bpl \Phi_A(\rho) \bpr =0$, confirming that the post-measurement state $\Phi_A(\rho)=\sum_ip_iA_i\otimes\rho_{\tx{B}|i}$ no longer has such resources. On the other hand, some quantumness still remains in the form of $B$-coherence and $B$-discord. Performing a measurement on part B yields the state $\Phi_{BA}(\rho)$ and the informational change
\be 
I \bpl \Phi_{BA}(\rho) \bpr -I \bpl \Phi_A(\rho) \bpr=-\big[C_B \bpl \vs{\Tr}{A}\Phi_A(\rho) \bpr +D_B \bpl \Phi_A(\rho) \bpr \big],
\label{step2}
\ee 
with $\vs{\Tr}{A}\Phi_A(\rho)=\vs{\rho}{B}$. The above expression shows that $B$-coherence and $B$-discord are removed upon the measurement of $B$, as expected. Noting that $D_A(\rho)+D_B \bpl \Phi_A(\rho) \bpr =D_{AB}(\rho)$, we sum Eqs. \eqref{step1} and \eqref{step2} to write the total resource suppressed so far as $I \bpl \Phi_{BA}(\rho) \bpr -I(\rho)=-[C_A(\vs{\rho}{A})+C_B(\vs{\rho}{B})+D_{AB}(\rho)]$. Clearly, $\{A,B\}$-related coherences and quantum correlations have been entirely removed. This motivates us to introduce the {\it quantumness} underlying the set $\mbb{O}\equiv\{A\otimes\vs{\mbb{1}}{B},\vs{\mbb{1}}{A}\otimes B\}$ (hereafter $\mbb{O}=\{A,B\}$, for short),
\be 
\mf{Q}_\mbb{O}(\rho)\coloneqq C_A(\vs{\rho}{A})+C_B(\vs{\rho}{B})+D_{AB}(\rho)=I(\rho)-I \bpl \Phi_{BA}(\rho) \bpr,
\label{Q_O}
\ee 
which can be interpreted as the amount of information removed from $\rho$ via $\{A,B\}$ measurements. The resulting state, $\Phi_{AB}(\rho)=\sum_{ij}p_{ij}A_i\otimes B_j$, still encodes an amount $I \bpl \Phi_{AB}(\rho) \bpr =\ms{H}(\{p_{ij}\})$ of information, where $\ms{H}(\{p_{ij}\})$ is the Shannon entropy of the distribution $p_{ij}$. This suggests that some quantum feature remains, meaning that the information in $\rho$ is not entirely encoded in the form of coherence and correlations. To appreciate this point, consider the set $\bar{\mbb{O}}\equiv\{\bar{A},\bar{B}\}$ of observables maximally noncommuting with $\mbb{O}=\{A,B\}$, that is, $[A,\bar{A}]\neq 0$ with their corresponding eigenbases satisfying $|\braket{a_i|\bar{a}_j}|^2=1/\vs{d}{A}$ (similarly for $B$ and $\bar{B}$). In this sense, $\mbb{O}$ and $\bar{\mbb{O}}$ are {\it maximally unbiased} (MU), in reference to the concept of maximally unbiased bases~\cite{durt2010}. It follows that $\Phi_{\bar{A}\bar{B}}\Phi_{AB}(\rho)=\mathbbm{1}/d$ and $I \bpl \Phi_{\bar{A}\bar{B}}\Phi_{AB}(\rho) \bpr =0$. That is, to erase all the information encoded in $\rho$, we still have to submit the system to measurements of the maximally noncommuting set $\bar{\mbb{O}}$. This analysis suggests that the remaining quantum element is {\em quantum incompatibility}, which we now briefly discuss.

Modern approaches have rephrased the notion of incompatibility, traditionally related to observables noncommutativity, as a quantum resource with operational interpretation and mathematical support~\cite{busch2013,beinosaari2015,toigo2018,toigo2019,guhne2019,cavalcanti2019,buscemi2020}. Recently, the two of us and a collaborator introduced the {\it context incompatibility}~\cite{martins2020}
\be 
\ms{I}_{\{\vs{\rho}{A},\vs{\mbb{O}}{A}\}}=I \bpl \Phi_{A_1}(\vs{\rho}{A}) \bpr -I \bpl \Phi_{A_2A_1}(\vs{\rho}{A}) \bpr,
\label{I_CA}
\ee 
such that the context is $\{\vs{\rho}{A},\vs{\mbb{O}}{A}\}\subset\mf{B}(\mc{H}_\tx{A})$, $A_{1,2}$ are observables acting on $\vs{\mc{H}}{A}$, and $\vs{\mbb{O}}{A}=\{A_1,A_2\}$. The above measure vanishes if and only if (i) $\Phi_{A_1}(\vs{\rho}{A})=\frac{\vs{\mbb{1}}{A}}{\vs{d}{A}}$ or (ii) $[A_1,A_2]=0$ $(\forall\,\vs{\rho}{A})$, and possesses an operational interpretation in a communication protocol. It naturally extends to bipartite scenarios, where the context is given by $\{\rho,\mbb{O}_1,\mbb{O}_2\}\subset \mf{B}\left(\vs{\mc{H}}{A}\otimes\vs{\mc{H}}{B}\right)$, with $\mbb{O}_1=\{A_1,B_1\}$, $\mbb{O}_2=\{A_2,B_2\}$, and $B_{1,2}\in\mf{B}(\vs{\mc{H}}{B})$:
\be 
\ms{I}_{\{\rho,\mbb{O}_1,\mbb{O}_2\}}=I \bpl \Phi_{\mbb{O}_1}(\rho) \bpr -I \bpl \Phi_{\mbb{O}_2\mbb{O}_1}(\rho) \bpr,
\label{I_C}
\ee 
where $\Phi_{\mbb{O}_k}\equiv\Phi_{A_kB_k}=\Phi_{A_k}\Phi_{B_k}$ and $\Phi_{\mbb{O}_2\mbb{O}_1}\equiv\Phi_{\mbb{O}_2}\Phi_{\mbb{O}_1}$. It can be demonstrated that the measure \eqref{I_C} has all the features that elects the original one [Eq.~\eqref{I_CA}] as a faithful quantifier of context incompatibility. 

Then, with $\bar{\mbb{O}}=\{\bar{A},\bar{B}\}$ being MU to $\mbb{O}=\{A,B\}$, we see that $\text{\small $\ms{I}_{\{\rho,\mbb{O},\bar{\mbb{O}}\}}$}=I \bpl \Phi_\mbb{O}(\rho) \bpr =I \bpl \Phi_{AB}(\rho) \bpr$, which is the precise amount of information that remained in the state in our previous discussion. This allows us to return to Eq.~\eqref{Q_O} to obtain
\be 
I(\rho)=\mf{Q}_\mbb{O}(\rho)+\ms{I}_{\{\rho,\mbb{O},\bar{\mbb{O}}\}}.
\label{I_QI}
\ee 
Note that the quantumness $\mf{Q}_\mbb{O}$ encompasses coherence and correlations associated with the set $\mbb{O}$, whereas $\text{\small $\ms{I}_{\{\rho,\mbb{O},\bar{\mbb{O}}\}}$}$, which is a nonclassical resource as well, is linked with both MU sets, $\mbb{O}$ and $\bar{\mbb{O}}$. Now, even though we can find, in general, more than one set $\bar{\mbb{O}}$ for each given context $\mbb{O}$, the choice of the former is constrained to its algebraic relation with the latter, so that the information decomposition can ultimately be related to $\mbb{O}$ solely. To see this, we note that for $\rho=\Phi_{\bar{\mbb{O}}}(\varrho)$ ($\forall\,\varrho$) one has $\text{\small $\ms{I}_{\{\Phi_{\bar{\mbb{O}}}(\varrho),\mbb{O},\bar{\mbb{O}}\}}$}=0$ and $I \bpl \Phi_{\bar{\mbb{O}}}(\varrho) \bpr =\mf{Q}_\mbb{O} \bpl \Phi_{\bar{\mbb{O}}}(\varrho) \bpr$. Then, referring to Eq.~\eqref{I_QI}, one has $\ms{I}_{\{\rho,\mbb{O},\bar{\mbb{O}}\}}=I \bpl \Phi_\mbb{O}(\rho)\bpr =\mf{Q}_{\bar{\mbb{O}}} \bpl \Phi_\mbb{O}(\rho) \bpr$, which naturally leads us to introduce the {\em incompatible quantumness} $\bar{\mf{Q}}$:
\be 
\ms{I}_{\{\rho,\mbb{O},\bar{\mbb{O}}\}}=\mf{Q}_{\bar{\mbb{O}}} \bpl \Phi_\mbb{O}(\rho) \bpr \eqqcolon \bar{\mf{Q}}_\mbb{O}(\rho).
\label{IQQtil}
\ee 
Finally, we arrive at the desired decomposition,
\be 
I(\rho)=\mf{Q}_\mbb{O}(\rho)+\bar{\mf{Q}}_\mbb{O}(\rho).
\label{I_expansion}
\ee 
In this compact form we can appreciate the quantum contents of information, namely, the quantumness $\mf{Q}_\mbb{O}$, encompassing quantum coherence and quantum correlations with respect to $\mbb{O}$, and the incompatible quantumness $\bar{\mf{Q}}_\mbb{O}$, which pinpoints the fundamental role of incompatibility for quantum information. As we have for the Lorentz invariant $\dr s^2=\dr\mathbf{r}^{\,2}-(c\dr t)^2$, expression \eqref{I_expansion} puts on the left-hand side the absolute quantity and, on the right-hand side, the frame-dependent objects, the ones that are accessed via measurements.

\section{Covariance}
The above discussion suggests that, while quantum coherence, quantum correlations, and quantum incompatibility are not absolute resources individually, they add up to an invariant one. However, to definitely prove information covariance, we need to concretely identify meaningful sets of observables and the unitary transformation $T$ that yields the jump from R's to A's perspective. Let us consider a scenario wherein two systems, A and B, are described by an observer R through $\mbb{C}\equiv\{\rho,\mbb{O}\}$, where $\mbb{O}=\{A,B\}$. The change to A's perspective, which reduces R and B to observed systems, cannot be implemented by effectively transforming both the observables and the state. In fact, there are two recipes that lead R's description, $\mbb{C}=\{\rho,\mbb{O}\}\subset\mf{B}(\mc{H})$, to A's description, $\mbb{C}'\equiv\{\rho',\mbb{O}'\}\subset\mf{B}(\mc{H}')$, namely,
\begin{subequations}
\begin{align}
&\text{Active Picture (AP)}:  &\{\rho',\mbb{O}'\}=\{T\rho T^\dag,\mbb{O}\},\\
&\text{Passive Picture (PP)}: &\{\rho',\mbb{O}'\}=\{\rho,T^\dag\mbb{O}T\},
\end{align}\label{pictures}
\end{subequations}
where $T:\mc{H}\mapsto\mc{H}'$, with $\mc{H}'=\vs{\mc{H}'}{R}\otimes\vs{\mc{H}'}{B}$ the Hilbert space adopted in A's perspective. The equivalence between these pictures comes by $\langle\mbb{O}'\rangle_{\rho'}=\Tr\big[(T\rho T^\dag)\mbb{O}\big]=\Tr\big[\rho(T^\dag\mbb{O}T)\big]$. It follows that $\Phi_\mbb{O}(T\rho T^\dag)=\Phi_{T^\dag \mbb{O} T}(\rho)$ (for any $\mbb{O}$) and $\mf{Q}_\mbb{O}(T\rho T^\dag)=\mf{Q}_{T^\dag\mbb{O}T}(\rho)$ (similarly for $\bar{\mf{Q}}$). Given Eq.~\eqref{I_expansion} and the invariance $I(\rho)=I(T\rho T^\dag)$ one guarantees that
\be 
\mf{Q}_\mbb{O}(\rho)+\bar{\mf{Q}}_\mbb{O}(\rho)=\mf{Q}_{\mbb{O}'}(\rho')+\bar{\mf{Q}}_{\mbb{O}'}(\rho')
\label{covariance}
\ee 
in both AP and PP. This formula, which constitutes the main result of this work, points the {\em form invariance} (covariance) of quantum information upon its measurement factorization in different reference frames, where distinct sets of observables, $\{\mbb{O},\bar{\mbb{O}}\}$ and $\{\mbb{O}',\bar{\mbb{O}}'\}$, are used for the actual access of information. In other words, it ensures that the total amount of quantum resources available is the same in all QRFs. It is very difficult to imagine a self-consistent way to decompose $I(\rho)$ in terms of measurement-based quantities of a purely classical information theory. This suggests that information covariance may be a fundamental principle of quantum mechanics. 

We now specialize our discussion to spatial coordinates, showing how to obtain the transformed context $\mbb{C}'$ from $\mbb{C}$. Let us discuss some aspects of general covariance starting with only two systems, R and A. Suppose that R prepares A in a position eigenstate $\ket{x}$. Because physics is deeply relational, there is no reason preventing us to believe that A has just prepared R in a state $\ket{-x}$. As far as momentum eigenstates are concerned, the statement ``R prepares A in $\ket{\vs{\mu}{\tiny AR} v}$'' is equivalent to ``A prepares R in $\ket{-\vs{\mu}{\tiny AR} v}$'', where $\vs{\mu}{\tiny AR}=\vs{m}{A}\vs{m}{R}/(\vs{m}{A}+\vs{m}{R})$ is the reduced mass of the system and $v$ is the relative speed. This sort of kinematical relationality is understood here as a fundamental premise that ought to be obeyed even when the mechanics is quantum and no matter how big or fast the involved bodies are (see Ref.~\cite{rovelli1996} for a related discussion). Although we have motivated our analysis in terms of position and momentum degrees of freedom, here we postulate that this must hold true for every physical state.

\vskip2mm
\noindent {\bf Postulate.} (No privileged quantum reference frame.)\\
{\em Every physical entity is allowed ``to observe'' or ``to be observed''. Therefore, whenever
{\em R} prepares {\em A} in a state $\ket{\psi}$, {\em A} automatically prepares {\em R} in a counterpart state $\ket{\psi'}$.}
\vskip2mm

\noindent For spatial coordinates, it is clear that $\psi'=-\psi$, with the unitary transformation being the usual parity operator, $T=\vs{\Pi}{A}$. The aforementioned preparation process in such two-particle universe is assumed to be consistent with all physical interactions and conservation laws~\cite{vedral2020}, and is not constrained to the absoluteness of time. If, for instance, R prepares A in a momentum superposition $\ket{-p}+\ket{p}$, then A prepares R in $\ket{p}+\ket{-p}$ \footnote{This postulate can be validated with a model that starts with an external reference frame which is posteriorly discarded~\cite{angelo2011,angelo2012}. The background independence prescribed by the postulate also requires that, when R prepares a spin-1/2 state $\ket{\uparrow}+\ket{\downarrow}$ for A, then R is automatically prepared in $\ket{\downarrow}+\ket{\uparrow}$ relative to A.}. It is important to note that this preparation is critically different from $\ket{-p}\ket{p}+\ket{p}\ket{-p}$, which is feasible only in the presence of a third system, say S, that can make sense of the motion of both A and R, and hence can prepare them in a momentum-conserving entangled state. In S's viewpoint, because of the existence of correlations, neither A or R is individually in a superposition of momentum states. In the two-particle universe, however, only a single degree of freedom exists (A's position relative to R, or vice-versa), so there is no ``informer'' to encode ``which-way information'' about the observed system. As a consequence, a fundamental wave-like behavior (a quantum superposition) manifests itself~\cite{ribeiro2015}.

Consider now a three-system scenario, with $\mbb{O}=\{\vs{X}{A},\vs{X}{B}\}$ being the set formed by the position operators of A and B relative to R. Here, we employ the transformation~\cite{giacomini2019,vanrietvelde2020} 
\be 
T=\vs{\Pi}{A}\,\E^{\I \vs{X}{A}\vs{P}{B}/\hbar},
\label{T_positions}
\ee 
where $\vs{\Pi}{A}$ is again the parity operator and $\E^{\I \vs{X}{A}\vs{P}{B}/\hbar}$ introduces to B a displacement conditioned to the position of A relative to R. (Throughout this work we use $\vs{X}{A}\vs{P}{B}$ as a shorthand for $\vs{X}{A}\otimes\vs{P}{B}$, and similarly for other products.) The new coordinate set then reads as $\mbb{O}'=T^{\dag}\mbb{O}T=\{\vs{X'}{R},\vs{X'}{B}\}\equiv\{-\vs{X}{A},\vs{X}{B}-\vs{X}{A}\}$, where $\vs{X'}{R}=T^\dag\vs{X}{A}T\in\mf{B}(\vs{\mc{H}'}{R})$ is the position of R relative to A and $\vs{X'}{B}=T^\dag\vs{X}{B}T\in\mf{B}(\vs{\mc{H}'}{B})$ is the position of B relative to A. The notation $\vs{X'}{R}=-\vs{X}{A}$ refers to an operator $\vs{X'}{R}$ that acts on $\vs{\mc{H}'}{R}$ respecting the same algebra with which $-\vs{X}{A}$ acts on $\vs{\mc{H}}{A}$. We have $\langle \vs{X'}{R}\rangle_{\rho'}=-\langle \vs{X}{A}\rangle_{\rho}$, thus confirming that $T$ promotes the system A to the role of reference frame. In terms of the AP, one finds the expected result $T\ket{u}\ket{v}=\ket{-u}\ket{v-u}$. To generalize Eq.~\eqref{T_positions} to a many-particle system one applies the replacement $\vs{P}{B}\to\sum_\tx{S}\vs{P}{S}$, with $\tx{S}=\tx{B},\tx{C},\tx{D}\cdots$. As for momenta, the transformation gives $T^\dag\{\vs{P}{A},\vs{P}{B}\}T=\{-\vs{P}{A}-\vs{P}{B},\vs{P}{B}\}\equiv\{\vs{P'}{R},\vs{P'}{B}\}$, which makes no link with the relative momenta (a mere consequence of the canonical Hamiltonian formalism~\cite{angelo2012,angelo2015}). The transformation that produces the correct relative momenta reads
\be 
\mc{T} \coloneqq \vs{\Pi}{A}\,\exp\text{\small $\left(\frac{\I\alpha\{\vs{X}{B},\vs{P}{B}\}}{2\hbar}\right)$}\exp\text{\small $\left(-\frac{\I}{\hbar}\frac{\vs{m}{B}}{\vs{m}{A}}\vs{X}{B}\vs{P}{A}\right)$},
\ee 
with $\alpha=\ln(\vs{\mu}{AB}/\vs{m}{B})$ and $\{\vs{X}{B},\vs{P}{B}\}\equiv\vs{X}{B}\vs{P}{B}+\vs{P}{B}\vs{X}{B}$, which gives $\mc{T}^\dag\{\vs{P}{A},\vs{P}{B}\}\mc{T}=\big\{-\vs{P}{A},\vs{\mu}{AB}\big(\tfrac{\vs{P}{B}}{\vs{m}{B}}-\tfrac{\vs{P}{A}}{\vs{m}{A}}\big)\big\}$. In this case, the new coordinates, $\mc{T}^\dag\{\vs{X}{A},\vs{X}{B}\}\mc{T}=\big\{-\big(\vs{X}{A}+\tfrac{\vs{m}{B}}{\vs{m}{A}}\vs{X}{B}\big),\tfrac{\mu}{\vs{m}{B}}\vs{X}{B}\big\}$, are not the desired relative positions.

We have seen, therefore, that positions and momenta explicitly admit unitary transformations switching the description to the viewpoint of the quantum particle A. Moreover, the whole treatment presented is such that R itself is a quantum system, so that no privileged external background needs to be conceived, thus ensuring the global relationality underlying Eq.~\eqref{covariance}. To illustrate the whole idea of our approach, we discuss in the next section two fundamental problems: (i) a framework whereby the universality of free fall is validated in the nonrelativistic quantum domain and (ii) a demonstration that elements of reality are observer-dependent. Also noteworthy is the fact that quantum mechanics admits a trivial Galilean counterpart of $\dr s^2$. To see this, consider the parts R, A, B, and C. In R's perspective, the position of C relative to B is computed as $\langle\vs{X}{C}-\vs{X}{B}\rangle_\rho$. The generalization of transformation \eqref{T_positions} to this case is $T=\vs{\Pi}{A}\exp\big[\I \vs{X}{A}(\vs{P}{B}+\vs{P}{C})/\hbar\big]$. By direct application of $T$ we find $\vs{X'}{C}=\vs{X}{C}-\vs{X}{A}$ and $\vs{X'}{B}=\vs{X}{B}-\vs{X}{A}$. It then follows that $\langle\vs{X'}{C}-\vs{X'}{B}\rangle_{\rho'}=\langle\vs{X}{C}-\vs{X}{B}\rangle_\rho$, which proves invariance in the Galilean space-time (see Ref.~\cite{angelo2015} for a discussion involving dynamics and Refs.~\cite{vanrietvelde2020,vanrietvelde2018} for related results in the perspective-neutral framework).

\section{Applications}
\subsection{Free fall in different reference frames}
\label{details}

We now use our results to discuss the so-called breakdown of the {\it universality of free fall} in the quantum framework~\cite{su1994,viola1997,williams2004,schlamminger2008,schlippert2014,seveso2017}, a problem that seems to preclude a pacific coexistence between quantum mechanics and the weak equivalence principle. Before starting, let us notice that the solution of the Schr\"odinger problem with initial state $\ket{\Psi_0}$ and Hamiltonian $H=P^2/(2m)+mgX$ describing a particle of mass $m$ under a uniform gravitational field $g$ can be written in the form
\be
\ket{\Psi_t}=\E^{-\I\,\Theta_t} \exp\left(-\frac{\I mgtX}{\hbar} \right) \exp \left(-\frac{\I P^2t}{2m\hbar} \right)\exp\left(\frac{\I gt^2P}{2\hbar} \right)\ket{\Psi_0},
\label{Psi_fall}
\ee
where $\Theta_t=mg^2t^3/(6\hbar)$. The two $P$-dependent unitary transformations have clear interpretations: one of them imposes a spatial shift $-gt^2/2$ on the wave function $\Psi_0(x)$ and the other introduces the typical free-evolution spreading. The others yield phases.

Let us now consider two particles, A and B, both in free fall motion from the perspective of an inertial observer R on Earth. Classically, if the particles depart from rest at heights $\vs{x}{A,B}$ such that $\vs{x}{B}-\vs{x}{A}=d$, then they fall under the same uniform gravitational acceleration $g$, the relative height $d$ does not change, and the times of flight do not depend on the masses of the particles. Quantum mechanically, however, R cannot simultaneously prepare well defined positions and momenta for the particles, so that the time-of-flight statistics are seen to depend on the masses $\vs{m}{A,B}$ of the falling particles. Associated with the typical spreading of moving wave packets, this result precludes the universality of free fall. Incidentally, one may still conceive an instance where the universality is restored, at least within the short-time approximation. In terms of the center-of-mass and relative canonical operators, 
\begin{subequations}
\begin{align} 
&\vs{X}{cm}=\text{\small $\frac{\vs{m}{A}\vs{X}{A}+\vs{m}{B}\vs{X}{B}}{M}$},&\quad &\vs{X}{r}=\vs{X}{B}-\vs{X}{A}, \\
&\vs{P}{cm}=\vs{P}{A}+\vs{P}{B}, &\quad &\vs{P}{r}=\mu\text{\small $\left(\frac{\vs{P}{B}}{\vs{m}{B}}-\frac{\vs{P}{A}}{\vs{m}{A}}\right)$},
\end{align}
\end{subequations}
where $M=\vs{m}{A}+\vs{m}{B}$ and $\mu=\vs{m}{A}\vs{m}{B}/M$, the original two-particle Hamiltonian $H$ is mapped onto $\utilde{H}$ as
\be 
H=\sum_{\tx{\tiny S}=\tx{\tiny A},\tx{\tiny B}}\left(\text{\small $\frac{\vs{P}{_S}^2}{2\vs{m}{_S}}$}+\vs{m}{_S}g\vs{X}{_S}\right)\,\,\stackrel{\utilde{T}}{\mapsto}\,\,\utilde{H}=\text{\small $\frac{\vs{P}{cm}^2}{2M}$}+Mg\vs{X}{cm}+\text{\small $\frac{\vs{P}{r}^2}{2\mu}$}.
\ee 
The unitary transformation $\utilde{T}$ is exhibited in Ref.~\cite{angelo2012}. It is clear that the center of mass is in free fall whereas the relative coordinate is in free motion. With the help of Eq.~\eqref{Psi_fall}, we propose, in terms of the $\{\vs{X}{cm},\vs{X}{r}\}$ eigenbases, the uncoupled solution
\be 
\ket{\utilde{\psi}_t}=\int\dr u\,\varphi_t \big(u-\tfrac{gt^2}{2}\big) \ket{u}\int\dr v\,\phi_t\big(v-d\big)\ket{v},
\ee 
such that $|\varphi_t|^2=\mc{G}_{\vs{\sigma}{cm}^t}(u-gt^2/2)$ and $|\phi_t|^2=\mc{G}_{\vs{\sigma}{r}^t}(v-d)$, with 
\be 
\mc{G}_\sigma(x-x_c)\coloneqq \frac{\exp\left[-\frac{(x-x_c)^2}{2\sigma^2}\right]}{\sqrt{2\pi\sigma^2}}.
\label{G}
\ee 
The function $\vs{\sigma}{s}^t=\vs{\sigma}{s}\big(1+t^2/\vs{t}{s}^2\big)^\text{\tiny $1/2$}$ gives the uncertainty associated with the degree of freedom $\tx{s}\in\{\tx{cm,r}\}$, where $\vs{t}{cm}=2M\vs{\sigma}{cm}^2/\hbar$ and $\vs{t}{r}=2\mu\vs{\sigma}{r}^2/\hbar$. This particular solution suggests how quantum mechanics restores the universality of free fall in the semiclassical limit: for small $\vs{\sigma}{r}$ and large $\vs{t}{r}$ (attainable for large masses and $\vs{m}{A}\gg\vs{m}{B}$), the $|\phi_t|^2$ spreading remains negligible for very long times, and thus the relative distance between the particles does not fluctuate.

We now consider the map $\ket{u}\ket{v}\mapsto \ket{u-\tfrac{\vs{m}{B}}{M}v}\ket{u+\tfrac{\vs{m}{A}}{M}v}$, which implements the transition from $\{\vs{x}{cm},\vs{x}{r}\}$ back to the coordinates $\{\vs{x}{A},\vs{x}{B}\}$ relative to R. With that, we find
\begin{subequations}
\beq 
&\displaystyle\ket{\psi_t}=\int\!\!\!\int\!\!\dr u \dr v\, \varphi_t\big(r_{u,v}-\chi_t\big)\,\phi_t\big(s_{u,v}-d\big) \ket{u}\ket{v},\quad & \\
&\displaystyle r_{u,v}=\text{\small $\frac{\vs{m}{A}u+\vs{m}{B}v}{M}$}, \qquad s_{u,v}=v-u,&
\eeq \label{psi_t}
\end{subequations}
where $\chi_t=gt^2/2$. To move to A's perspective, we apply $T$ as given by Eq. (16), which yields $\vs{x'}{R}=-\vs{x}{A}$ and $\vs{x'}{B}=\vs{x}{B}-\vs{x}{A}$. We obtain $T\ket{u}\ket{v}=\ket{-u}\ket{v-u}$ and
\begin{subequations}
\beq 
&\displaystyle\ket{\psi_t'}=\int\!\!\!\int\!\!\dr u \dr v\,\varphi_t\left(r_{u,v}'-\chi_t\right)\,\phi_t(s_{u,v}'-d) \ket{u}\ket{v},\quad &\\
&\displaystyle r_{u,v}'=r_{-u,v-u}, \qquad s_{u,v}'=s_{-u,v-u}.&
\eeq \label{psi_t'}
\end{subequations}
To compute the quantumenesses of the states \eqref{psi_t} and \eqref{psi_t'}, we adopt the discretization formalism developed in Ref.~\cite{freire2019}. This amounts to setting $(u,v)=(i,j)\delta q$, with $\delta q$ the spatial resolution (which is the same in every reference frame), and replacing the integrals with summations running over integers $i,j\in[-L,L]$, where $L=(\xi-1)/2$ and $\xi=2\pi\hbar/(\delta q\delta p)$ (space dimension). Here, $\delta p$ denotes the momentum resolution. In addition, projectors are given by $\Pi_i=\delta q\,\ket{i}\bra{i}$, with $\Pi_i\Pi_{i'}=\delta_{ii'}\Pi_i$ and $\sum_{i=-L}^L\Pi_i=\mbb{1}$, $\braket{i|i'}=\delta_{ii'}/\delta q$, and $\braket{j|j'}=\delta_{jj'}/\delta q$. Then, we have
\begin{subequations}
\begin{align} 
&\ket{\psi_t}=\sum_{i,j}\delta q\,\ov{\varphi}_t\big(\ov{r}_{i,j}-\ov{\chi}_t\big)\,\ov{\phi}_t\big(\ov{s}_{i,j}-\ov{d}\big)\ket{i}\ket{j}, \\
&\ket{\psi_t'}=\sum_{i,j}\delta q\,\ov{\varphi}_t\big(\ov{r}_{i,j}'-\ov{\chi}_t\big)\,\ov{\phi}_t\big(\ov{s}_{i,j}'-\ov{d}\big)\ket{i}\ket{j},
\end{align} \label{psi_discrete}
\end{subequations}
where $(\ov{\varphi}_t,\ov{\phi}_t)\equiv (\varphi_t,\phi_t)/\sqrt{\delta q}$ are dimensionless amplitudes such that $|\ov{\varphi}_t|^2=\mc{G}_{\vs{\ov{\sigma}}{cm}^t}\big(\ov{r}_{i,j}-\ov{\chi}_t\big)$ and $|\ov{\phi}_t|^2=\mc{G}_{\vs{\ov{\sigma}}{r}^t}\big(\ov{s}_{i,j}-\ov{d}\big)$. The set $\{\ov{r}_{i,j},\ov{s}_{i,j},\ov{\chi}_t,\ov{d},\vs{\ov{\sigma}}{cm}^t,\vs{\ov{\sigma}}{r}^t\}$ are formed with quantities normalized with $\delta q$.

\begin{figure}[htb]
\centerline{\includegraphics[scale=0.55]{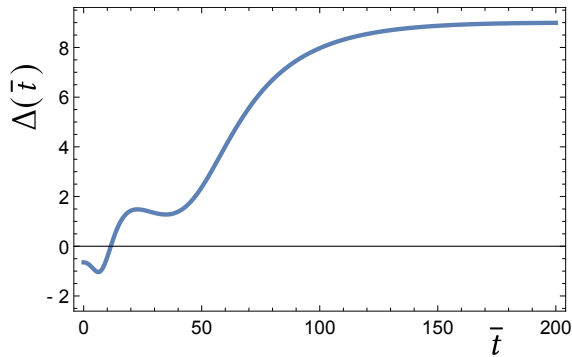}}
\caption{\small Percentage difference $\Delta (\,\tx{\small $\ov{t}$}\,)$, with respect to $I(\rho_0)$, between the quantumnesses $\mf{Q}_{\vs{X}{A}\vs{X}{B}}(\rho_t)$, in R's frame, and $\mf{Q}_{\vs{X'}{A}\vs{X'}{B}}(\rho_t')$, in A's frame, as a function of the scaled time $\ov{t}=t/\tau$. In this simulation, we have used $\tau\simeq 10^{-10}$ s, $\vs{m}{A(B)}\simeq 2.4\times 10^{-10}$ kg, $\tx{\small $\ov{d}$}=3$, $\vs{\ov{\sigma}}{cm}=7$, and $\vs{\ov{\sigma}}{r}=3$. The discretization scheme is characterized by $L=15$, implying a space dimension $\vs{d}{A(B)}=\xi=31$, $\delta q\,\delta p\cong\hbar/5$, and $\delta q\sim\delta p\sim 0.45\tx{\small $\sqrt{\hbar}$}$ (with SI units).}
\label{fig1}
\end{figure}

Now, in the regime where $\eta\equiv\frac{\vs{m}{B}}{\vs{m}{A}}\to 0$, one readily sees that $(\ov{r}_{i,j},\ov{s}_{i,j})=(i,j-i)$ and $(\ov{r}_{i,j}',\ov{s}_{i,j}')=(-i,j)$, which renders $\rho_t'=\ket{\psi_t'}\bra{\psi_t'}$ separable. This proves that
\be 
0< D_{\vs{X}{A}\vs{X}{B}}(\rho_t)\neq D_{\vs{X'}{A}\vs{X'}{B}}(\rho_t')=0,
\ee 
illustrating that quantum correlations are not generally preserved under changes of quantum references frames. With respect to the state $\vs{\rho}{A}=\vs{\Tr}{B}(\rho_t)$, where $\rho_t=\ket{\psi_t}\bra{\psi_t}$, we introduce $f(i)\equiv\ov{\varphi}_t(i-\ov{\chi}_t)$ and $g(j-i)\equiv\ov{\phi}_t(j-i-\ov{d})$ to compactly write
\be 
\vs{\rho}{A}=\sum_{ii'}\delta q^2\,f(i)\,f^*(i')\,\gamma_{ii'}\ket{i}\bra{i'},
\ee
where the parameter $\gamma_{ii'}=\sum_j g(j-i)\,g^*(j-i')$, which depends on $\vs{\sigma}{r}$, regulates the quantum coherence of $\vs{\rho}{A}$ but plays no role for $\vs{\rho}{A}'$. In particular, for $\vs{\sigma}{r}\to 0$, we have $\gamma_{ii'}=\delta_{ii'}$ and
\be 
0=C_{\vs{X}{A}}(\vs{\rho}{A})\neq C_{\vs{X'}{A}}(\vs{\rho}{A}')>0.
\ee 
This shows that quantum coherence is not an invariant resource either. Returning to Eqs.~\eqref{psi_discrete}, one analytically finds, via direct calculations, $\Phi_{\vs{X}{A}\vs{X}{B}}(\rho_t)=\sum_{i,j}\delta q^2\,\wp_{ij}\ket{i,j}\bra{i,j}$ and $\Phi_{\vs{X'}{A}\vs{X'}{B}}(\rho_t')=\sum_{i,j}\delta q^2\,\wp'_{ij}\ket{i,j}\bra{i,j}$, where
\begin{subequations}
\begin{align}
&\wp_{ij}=\mc{G}_{\vs{\ov{\sigma}}{cm}^t}\big(\ov{r}_{i,j}-\ov{\chi}_t\big)\,\mc{G}_{\vs{\ov{\sigma}}{r}^t}\big(\ov{s}_{i,j}-\ov{d}\big), \\
&\wp'_{ij}=\mc{G}_{\vs{\ov{\sigma}}{cm}^t}\big(\ov{r}_{i,j}'-\ov{\chi}_t\big)\,\mc{G}_{\vs{\ov{\sigma}}{r}^t}\big(\ov{s}_{i,j}'-\ov{d}\big).
\end{align}\label{pp'}
\end{subequations}
With these expressions, which hold for arbitrary $\eta$, we obtain, via Eq. (8), $\mf{Q}_{\vs{X}{A}\vs{X}{B}}(\rho_t)=S(\Phi_{\vs{X}{A}\vs{X}{B}}(\rho_t))=\ms{H}(\{\wp_{ij}\})$ and its counterpart $\mf{Q}_{\vs{X'}{A}\vs{X'}{B}}(\rho_t')=\ms{H}(\{\wp'_{ij}\})$ in A's frame. It can be readily inferred that, in general, $\wp_{ij}\neq\wp'_{ij}$ and
\be 
\mf{Q}_{\vs{X}{A}\vs{X}{B}}(\rho_t)\neq \mf{Q}_{\vs{X'}{A}\vs{X'}{B}}(\rho_t'),
\ee 
meaning that not even quantumness is invariant. In what follows we present the results of a simulation for the case involving equal-mass particles. The discretized model adopted here is such that $L=15$, implying the space dimension $\vs{d}{A(B)}=\xi=31$ and $\delta q\,\delta p\cong\hbar/5$ (roughly, $\delta q\sim\delta p\sim 0.45\tx{\small $\sqrt{\hbar}$}$, with pertinent SI units). Here, the time evolution is analyzed in terms of the dimensionless time $\ov{t}=t/\tau$, where $\tau\equiv 2\vs{m}{A}\delta q^2/\hbar\simeq 10^{-10}$ s with $\vs{m}{A(B)}\simeq 2.4\times 10^{-10}$ kg. Also, we use $\tx{\small $\ov{d}$}=3$, $\vs{\ov{\sigma}}{cm}=7$, and $\vs{\ov{\sigma}}{r}=3$. It is worth mentioning that, to ensure the physical validity of the discretized model, the probability distributions $\wp_{ij}$ and $\wp_{ij}'$ were monitored and, when necessary, suitably renormalized for all times of the simulation. Figure~\ref{fig1} shows the behavior of the quantity
\be 
\Delta (\,\tx{\small $\ov{t}$}\,)\coloneqq \frac{\mf{Q}_{\vs{X}{A}\vs{X}{B}}(\rho_t)-\mf{Q}_{\vs{X'}{A}\vs{X'}{B}}(\rho_t')}{I(\rho_0)}\,\times 100\%,
\ee 
where $I(\rho_0)=\ln{\xi^2}$. It gives the percentage difference, with respect to the invariant information $I(\rho_0)$, between the quantumnesses available to the quantum reference frames R and A. This result illustrates that quantum coherence and quantum correlations do not form an invariant; to this end, the incompatible quantumness $\bar{\mf{Q}}_{\vs{X}{A}\vs{X}{B}}(\rho_t)=I(\rho_0)-\mf{Q}_{\vs{X}{A}\vs{X}{B}}(\rho_t)$ is an indispensable parcel. 
%

\subsection{Relativity of reality}

Recently, a criterion of physical reality was introduced~\cite{bilobran2015} which has been shown to be rather enlightening with respect to both foundational and applied issues such as (i) the discovery of an information-reality complementarity~\cite{dieguez2018}, (ii) the definition of bipartite~\cite{gomes2018,gomes2019} and tripartite~\cite{fucci2019} aspects of nonlocality that are fundamentally different from those deriving from Bell inequality violations, (iii) the discussion of realism violations and nonlocality in a two-walker system~\cite{orthey2019}, and (iv) the proposition of an alternative solution to Hardy's paradox~\cite{engelbert2019}. The key premise behind the aforementioned criterion is that after a measurement is conducted of an observable $A\in\mf{B}(\vs{\mc{H}}{A})$, there must be an element of reality associated with $A$ (or, $A$ is an element of reality), even when the measurement outcome is kept in secret. In this context, the unrevealed-measurement state $\Phi_A(\rho)$ can thus be taken as a state of reality for $A$ (henceforth referred to as an $A$-reality state). Accordingly, if $\Omega=\Phi_A(\rho)$, then $\Phi_A(\Omega)=\Omega$, meaning that further unrevealed measurements of $A$ on a $A$-reality state $\Omega$ does not alter the fact that $A$ is already an element of reality. It then follows that $\rho=\Phi_A(\rho)$ can be adopted as a criterion of $A$-reality and, as a consequence,
\be 
\mf{I}_A(\rho)\coloneqq S\big(\Phi_A(\rho)\big)-S(\rho)
\label{frakI}
\ee 
emerges as a quantifier of the degree with which the $A$-reality criterion is violated for a given $\rho$. Called {\em irreality}, $\mathfrak{I}_A(\rho)$ is nonnegative and vanishes if and only if $\rho=\Phi_A(\rho)$. Interestingly, it is easy to verify that $\mf{I}_A(\rho)=C_A(\rho)+D_A(\rho)$, implying that $\mf{Q}_{\mbb{O}}(\rho)=\mf{I}_A(\rho)+\mf{I}_B\big(\Phi_A(\rho)\big)$. This shows that the notion of quantumness and its incompatible counterpart $\bar{\mf{Q}}$ can be entirely rephrased in terms of irreality.

Now, via the AP-PP equivalence one can easily check that $S\big(\Phi_A(T\rho T^\dag)\big)$ is equal to $S \bpl \Phi_{T^\dag A T}(\rho) \bpr$, but none of these versions can be equated to $S\big(\Phi_A(\rho)\big)$. This is to say that $\mf{I}_A(T\rho T^\dag)\neq \mf{I}_A(\rho)$, which implies that quantum irrealism as diagnosed via Eq.~\eqref{frakI} is not absolute. A simple illustration of this fact can be given by use of the discrete approach (see Ref.~\cite{freire2019} for further details), within which a position eigenstate is written as $\ket{x_k}=\ket{k\,\delta q}$, where $k\in\mbb{Z}$ and $\delta q$ is the experimental resolution for a position measurement. Suppose that R prepares the state
\be 
\ket{\psi}=\delta q\left(\text{\small $\frac{\ket{i}\ket{j}+\ket{i+k}\ket{j+k}}{\sqrt{2}}$}\right),
\ee 
where $i,j,k\in\mbb{Z}^*$. Since $\braket{i|i'}=\delta_{i,j}/\delta q$, one has $\braket{\psi|\psi}~\!\!\!=~\!1\!$. For concreteness, we can imagine the situation where a diatomic molecule, with atoms A and B, has just passed through a double-slit setup, with $k\delta q$ being the spatial separation of the slits. Using the projectors $\Pi_i=\delta q\,\ket{i}\bra{i}$ for both subspaces, one can apply standard techniques of discrete spaces algebra to obtain $\vs{\rho}{B}=(\ket{j}\bra{j}+\ket{j+k}\bra{j+k})\text{\small $\frac{\delta q}{2}$}=~\!\Phi_{\vs{X}{B}}(\vs{\rho}{B})$, $\Phi_{\vs{X}{B}}(\rho)=(\ket{i,j}\bra{i,j}+\ket{i+k,j+k}\bra{i+k,j+k})\text{\small $\frac{\delta q}{2}$}$, and $\Phi_{\vs{X}{A}\vs{X}{B}}(\rho)=\Phi_{\vs{X}{B}}(\rho)$. From direct calculations it follows that $C_{\vs{X}{B}}(\vs{\rho}{B})=0$ and $\mf{I}_{\vs{X}{B}}(\rho)=D_{\vs{X}{B}}(\rho)=\ln{2}$. Hence, from R's perspective, B's position is not an element of reality. On the other hand, form A's perspective we have
\be 
\ket{\psi'}=T\ket{\psi}=\delta q\left(\text{\small $\frac{\ket{-i}+\ket{-i-k}}{\sqrt{2}}$}\right)\ket{j-i}.
\ee 
We see that B's position relative to A is now well defined (atom A always ``sees'' atom B alongside). It is not difficult to show that $\mf{I}_{\vs{X'}{B}}(\rho')=C_{\vs{X'}{B}}(\rho')=D_{\vs{X'}{B}}(\rho')=0$, which confirms that B's position is an element of reality for A. Therefore, R and A do not agree on the degree of realism underlying B's position. As a side remark, we note that, incidentally, in this example we have $\mf{Q}_{\vs{X}{A}\vs{X}{B}}(\rho)=\mf{Q}_{\vs{X'}{A}\vs{X'}{B}}(\rho')$ (see Sec. \ref{details} for a counter-example). 

Before closing this section, it is useful to point out that the map $\ket{a}\ket{b}\mapsto\ket{-a}\ket{b-a}$, which trivially follows from the application of the transformation \eqref{T_positions} to position eigenstates, does not directly apply to the centers of Gaussian states. In fact, as we show in the Appendix \ref{IntEnt}, there is always a residual entanglement in the transformed state. 

\section{Discussion}
Under the premise of general covariance, significant efforts have recently been made toward the development of physical models that encompass QRFs and the transformations among them. The theoretical implications of such move are ubiquitous, ranging from quantum information and thermodynamics to quantum gravity and cosmology. Here, we have shown that, for symmetry groups generated by unitary transformations, the total amount of quantum resources, as given by the sum of coherences, correlations, and incompatibility, is the same for all QRFs. This means that while two distinct QRFs may not agree upon the amount of some specific resource, they will never disagree on the total amount. Our result indicates quantum resources covariance as a relevant ingredient for deeply relational models of nature.

\section*{Acknowledgments} 
We thank Philipp H\"ohn for very useful comments on an earlier version of this work 
and  acknowledge the Brazilian funding agency CNPq, under the grants 160986/2017-6 (M.F.S.) and 303111/2017-8 (R.M.A.), and the National Institute for Science and Technology of Quantum Information (CNPq, INCT-IQ 465469/2014-0).

\appendix
\section{Intrinsic entanglement}
\label{IntEnt}

There is an important aspect that is sometimes overlooked with respect to the transformation \eqref{T_positions}. Here we show that the map $\ket{a}\ket{b}\stackrel{\text{\tiny $T$}}{\mapsto}\ket{-a}\ket{b-a}$ does not directly apply to the centers of Gaussian states, not even for very sharp ones. Consider the state
\be 
\ket{\psi}=\int\dr u\dr v\,\mc{G}_{\Delta}^\tx{\tiny $1/2$}(u-a)\,\mc{G}_{\delta}^\tx{\tiny $1/2$}(v-b)\ket{u}\ket{v}
\label{product}
\ee 
describing the physics of systems A and B relative to R, with $\mc{G}_\sigma$ given by Eq.~\eqref{G}. Using the map given above and performing a change of dummy variables, we find 
\be 
\ket{\psi'}=\int\dr u\dr v  \,\mc{G}_{\Delta}^\tx{\tiny $1/2$}(-u-a)\,\mc{G}_{\delta}^\tx{\tiny $1/2$}(v-u-b)\ket{u}\ket{v},
\label{entangled}
\ee 
which gives the physics relative to A. The product of Gaussian functions in the integrand is proportional to
\be 
\exp\left(-\frac{(u+\alpha)^2}{4\zeta^2} \right) \ \exp\left(-\frac{(v-b)^2}{4\delta^2} \right) \ \exp\left(\frac{u(v-b)}{2\delta^2} \right),
\ee 
with $\alpha=a (\zeta/\Delta)^2$ and $\zeta=\delta\,\Delta/\sqrt{\delta^2+\Delta^2}$. In contrast with the case where center-of-mass and relative coordinates are used \cite{angelo2011}, the above transformation does not yield any reasonable regime where the crossing term can be neglected. Thus, the expansion \eqref{product}, which might be denoted $\ket{\psi}=\ket{a}\ket{b}$ in reference to a product of sharp states centered at $a$ and $b$, is not mapped onto $\ket{\psi'}=\ket{-a}\ket{b-a}$, not even approximately, because $\ket{\psi'}$ is strongly entangled. That is, the transformation rules for position eigenstates do not trivially apply to sharp Gaussian states.



\begin{thebibliography}{99}

\bibitem{aharonov1984} Y. Aharonov and T. Kaufherr, Phys. Rev. D {\bf 30}, 368 (1984).

\bibitem{poulin2007} D. Poulin and J. Yard, New J. Phys. {\bf 9}, 156
 (2007).

\bibitem{angelo2011} R. M. Angelo, N. Brunner, S. Popescu, A. Short, and P. Skrzypczyk, J. Phys. A: Math. Theor. {\bf 44}, 145304 (2011).

\bibitem{angelo2012} R. M. Angelo and A. D. Ribeiro, J. Phys. A: Math. Theor. {\bf 45}, 465306 (2012).

\bibitem{angelo2015} S. T. Pereira and R. M. Angelo, Phys. Rev. A {\bf 91}, 022107 (2015).

\bibitem{bartlett2006} S. D. Bartlett, T. Rudolph, R. W. Spekkens, and P. S. Turner, New J. Phys. {\bf 8}, 58 (2006).

\bibitem{bartlett2007} S. D. Bartlett, T. Rudolph, and R. W. Spekkens, Rev. Mod. Phys. {\bf 79}, 555 (2007).

\bibitem{massar1995} S. Massar and S. Popescu, Phys. Rev. Lett. {\bf 74}, 1259 (1995).

\bibitem{bartlett2009} S. D. Bartlett, T. Rudolph, R. W. Spekkens, and P. S. Turner, New J. Phys. {\bf 11}, 063013 (2009).

\bibitem{costa2009} F. Costa, N. Harrigan, T. Rudolph, and \v{C}. Brukner, New J. Phys. {\bf 11}, 123007 (2009).

\bibitem{liang2010} Y.-C. Liang, N. Harrigan, S. D. Bartlett, and T. Rudolph, Phys. Rev. Lett. {\bf 104}, 050401 (2010).

\bibitem{gour2008} G. Gour and R. W. Spekkens, New J. Phys. {\bf 10}, 033023 (2008).

\bibitem{popescu2018} S. Popescu, A. B. Sainz, A. J. Short, and A. Winter, Phil. Trans. Roy. Soc. A {\bf 376}, 20180111 (2018).

\bibitem{peres2002} A. Peres, P. F. Scudo, and D. R. Terno, Phys. Rev. Lett. {\bf 88}, 230402 (2002).

\bibitem{gingrich2002} R. M. Gingrich and C. Adami, Phys. Rev. Lett. {\bf 89}, 270402 (2002).

\bibitem{peres2004} A. Peres and D. R. Terno, Rev. Mod. Phys. {\bf 76}, 93 (2004).

\bibitem{giacomini22019} F. Giacomini, E. Castro-Ruiz, and \v{C}. Brukner, Phys. Rev. Lett. {\bf 123}, 090404 (2019).

\bibitem{rovelli1991-1} C. Rovelli, Classical Quantum Gravity {\bf 8}, 297 (1991). 

\bibitem{rovelli1991-2} C. Rovelli, Classical Quantum Gravity {\bf 8}, 317 (1991).

\bibitem{dittrich2006} B. Dittrich, Classical Quantum Gravity {\bf 23} 6155 (2006).

\bibitem{dittrich2007} B. Dittrich, Gen. Relativ. Gravit. {\bf 39}, 1891 (2007). 

\bibitem{girelli2008} F. Girelli and D. Poulin, Phys. Rev. D {\bf 77}, 104012 (2008).

\bibitem{hohn2019} P. A. H\"ohn, Universe {\bf 5}, 116 (2019).

\bibitem{vanrietvelde2020} A. Vanrietvelde, P. A. H\"{o}hn, F. Giacomini, and E. Castro-Ruiz, Quantum {\bf 4}, 225 (2020).

\bibitem{ruiz2020} E. Castro-Ruiz, F. Giacomini, A. Belenchia, and \v{C}. Brukner, Nat. Comm. {\bf 11}, 2672 (2020).

\bibitem{bojowald2011-1} M. Bojowald, P. A. H\"ohn, and A. Tsobanjan, Classical Quantum Gravity {\bf 28}, 035006 (2011).

\bibitem{bojowald2011-2} M. Bojowald, P. A. H\"ohn, and A. Tsobanjan, Phys. Rev. D {\bf 83}, 125023 (2011).

\bibitem{hohn2012} P. A. H\"ohn, E. Kubalova, and A. Tsobanjan, Phys. Rev. D {\bf 86}, 065014 (2012).

\bibitem{henderson2020} L. J. Henderson, A. Belenchia, E. Castro-Ruiz, C. Budroni, M. Zych, \v{C}. Brukner, and R. B. Mann, Phys. Rev. Lett. {\bf 125}, 131602 (2020).

\bibitem{barbado2020} L. C. Barbado, E. Castro-Ruiz, L. Apadula, and \v{C}. Brukner, Phys. Rev. D {\bf 102}, 045002 (2020).

\bibitem{lan1961} R. Landauer, IBM J. Research and Develop. {\bf 5}, 183 (1961).

\bibitem{ben1987} C. H. Bennet, Sci. Am. {\bf 257}, 108 (1987).

\bibitem{acosta2020} A. C. S. Costa and R. M. Angelo, Quantum Inf. Process. {\bf 19}, 325 (2020). 

\bibitem{nielsen2000} M. A. Nielsen and I. L. Chuang, {\it Quantum Computation and Quantum Information} (Cambridge University Press, Cambridge, UK, 2000).

\bibitem{shenoy2017} A. Shenoy-Hejamadi, A. Anirban Pathak, and R. Radhakrishna, Quanta {\bf 6}, 1 (2017).

\bibitem{anders2017} J. Anders and M. Esposito, New J. Phys. {\bf 19}, 010201 (2017).

\bibitem{dakic2011} B. Dakic and \v{C}. Brukner, Quantum Theory and Beyond: Is Entanglement Special? in {\it Deep Beauty: Understanding the Quantum World through Mathematical Innovation}, Ed. H. Halvorson (Cambridge University Press, 2011), 365-392.

\bibitem{masanes2011} L. Masanes and M. P. M\"uller, New J. Phys. {\bf 1}3, 063001 (2011).

\bibitem{chiribella2011} G. Chiribella, G. M. D'Ariano and P. Perinotti, Phys. Rev. A {\bf 84}, 012311 (2011).

\bibitem{hohn2017-Q} P. A. H\"ohn, Quantum {\bf 1}, 38 (2017).

\bibitem{hohn2017-PRA} P. A. H\"ohn and C. S. P. Wever, Phys. Rev. A {\bf 95}, 012102 (2017).

\bibitem{gour2015} G. Gour, M. P. M\"uller, V. Narasimhachar, R. W. Spekkens, and N. Y. Halpern, Phys. Rep. {\bf 583}, 1-58 (2015).

\bibitem{horodecki2003} M. Horodecki, P. Horodecki, and J. Oppenheim, Phys. Rev. A {\bf 67}, 062104 (2003).

\bibitem{vanrietvelde2018} A. Vanrietvelde, P. A. H\"ohn, and F. Giacomini, arXiv:1809.05093.

\bibitem{anne2020} A.-C. de la Hamette and T. D. Galley, Quantum {\bf 4}, 367 (2020).

\bibitem{chitambar2019} E. Chitambar and G. Gour, Rev. Mod. Phys. {\bf 91}, 025001 (2019).

\bibitem{cerf1997} N. J. Cerf and C. Adami, Phys. Rev. Lett. {\bf 79}, 5194 (1997).

\bibitem{horodecki2005} M. Horodecki, J. Oppenheim, and A. Winter, Nature {\bf 436}, 673 (2005).

\bibitem{horodecki2007} M. Horodecki, J. Oppenheim, and A. Winter, Comm. Math. Phys. {\bf 269}, 107 (2007).

\bibitem{horodecki2009} R. Horodecki, P. Horodecki, M. Horodecki, and K. Horodecki, Rev. Mod. Phys. {\bf 81}, 865 (2009).

\bibitem{baumgratz2014} T. Baumgratz, M. Cramer, and M. B. Plenio, Phys. Rev. Lett. {\bf 113}, 140401 (2014).

\bibitem{girolami2014} D. Girolami, Phys. Rev. Lett. {\bf 113} 170401 (2014).

\bibitem{streltsov2015} A. Streltsov, U. Singh, H. S. Dhar, M. N. Bera, and G. Adesso, Phys. Rev. Lett. {\bf 115}, 020403 (2015).

\bibitem{streltsov2017} A. Streltsov, G. Adesso, and M. B. Plenio, Rev. Mod. Phys. {\bf 89}, 041003 (2017).

\bibitem{giacomini2019} F. Giacomini, E. Castro-Ruiz, and \v{C}. Brukner, Nat. Commun., {\bf 10}, 494 (2019).

\bibitem{ollivier2001} H. Ollivier and W. H. Zurek, Phys. Rev. Lett. {\bf 88}, 017901 (2001).

\bibitem{henderson2001} L. Henderson and V. Vedral, J. Phys. A: Math. Gen. {\bf 34}, 6899 (2001).

\bibitem{rulli2011} C. C. Rulli and M. S. Sarandy, Phys. Rev. A {\bf 84}, 042109 (2011).

\bibitem{durt2010} T. Durt, B.-G. Englert, I. Bengtsson, and K. \.{Z}yczkowski, Int. J. Quantum Info. {\bf 8}, 535 (2010).

\bibitem{busch2013} P. Busch, T. Heinosaari, J. Schultz, and N. Stevens, Europhys. Lett. {\bf 103} 10002 (2013).

\bibitem{beinosaari2015} T. Heinosaari, J. Kiukas, and D. Reitzner, Phys. Rev. A {\bf 92}, 022115 (2015).

\bibitem{toigo2018} C. Carmeli, T. Heinosaari, and A. Toigo, Phys. Rev. A {\bf 98}, 012126 (2018).

\bibitem{toigo2019} C. Carmeli, T. Heinosaari, and A. Toigo, Phys. Rev. Lett. {\bf 122}, 130402 (2019).

\bibitem{guhne2019} R. Uola, T. Kraft, J. Shang, X.-D. Yu, and O. G\"uhne, Phys. Rev. Lett. {\bf 122}, 130404 (2019).

\bibitem{cavalcanti2019} P. Skrzypczyk, I. \v{S}upi\'{c}, and D. Cavalcanti, Phys. Rev. Lett. {\bf 122}, 130403 (2019).

\bibitem{buscemi2020} F. Buscemi, E. Chitambar, and W. Zhou, Phys. Rev. Lett. {\bf 124}, 120401 (2020).

\bibitem{martins2020} E. Martins, M. F. Savi, and R. M. Angelo, Phys. Rev. A {\bf 102}, 050201(R) (2020).

\bibitem{rovelli1996} C. Rovelli, Int. J. Theor. Phys. {\bf 35}, 1637 (1996).

\bibitem{vedral2020} C. Marletto and V. Vedral, arXiv:2005.00138.

\bibitem{ribeiro2015} R. M. Angelo and A. D. Ribeiro, Found. Phys. {\bf 45}, 1407 (2015).

\bibitem{su1994} Y. Su, B. R. Heckel, E. G. Adelberger, J. H. Gundlach, M. Harris, G. L. Smith, and H. E. Swanson, Phys. Rev. D {\bf 50}, 3614 (1994).

\bibitem{viola1997} L. Viola and R. Onofrio, Phys. Rev. D {\bf 55}, 455 (1997).

\bibitem{williams2004} J. G. Williams, S. G. Turyshev, and D. H. Boggs, Phys. Rev. Lett. {\bf 93}, 261101 (2004).

\bibitem{schlamminger2008} S. Schlamminger, K.-Y. Choi, T. A. Wagner, J. H. Gundlach, and E. G. Adelberger, Phys. Rev. Lett. {\bf 100}, 041101 (2008).

\bibitem{schlippert2014} D. Schlippert, J. Hartwig, H. Albers, L. L. Richardson, C. Schubert, A. Roura, W. P. Schleich, W. Ertmer, and E. M. Rasel, Phys. Rev. Lett. {\bf 112}, 203002 (2014).

\bibitem{seveso2017} L. Seveso and M. Paris, Ann. Phys. {\bf 380}, 213 (2017).

\bibitem{freire2019} I. S. Freire and R. M. Angelo, Phys. Rev. A {\bf 100}, 022105 (2019).

\bibitem{bilobran2015} A. L. O. Bilobran and R. M. Angelo, Europhys. Lett. {\bf 112}, 40005 (2015).

\bibitem{dieguez2018} P. R. Dieguez and R. M. Angelo, Phys. Rev. A {\bf 97}, 022107 (2018).

\bibitem{gomes2018} V. S. Gomes and R. M. Angelo, Phys. Rev. A {\bf 97}, 012123 (2018).

\bibitem{gomes2019} V. S. Gomes and R. M. Angelo, Phys. Rev. A {\bf 99}, 012109 (2019).

\bibitem{fucci2019} D. M. Fucci and R. M. Angelo, Phys. Rev. A {\bf 100}, 062101 (2019).

\bibitem{orthey2019} A. C. Orthey Jr. and R. M. Angelo, Phys. Rev. A {\bf 100}, 042110 (2019).

\bibitem{engelbert2019} N. G. Engelbert and R. M. Angelo, Found. Phys. {\bf 50}, 105 (2020).


\end{thebibliography}
\end{document}